\documentclass{aipproc}
\usepackage{epsfig}

\layoutstyle{6x9} 

\begin{document}


\title{Saturation properties of nuclear matter in a relativistic mean field model constrained by quark dynamics}

\author{R. Huguet}{address={Centre d'Etudes Nucl\'{e}aires de Bordeaux-Gradignan, CNRS-IN2P3 \\ Universit\'{e} Bordeaux 1, Le Haut-Vigneau, 33170 Gradignan Cedex, France}, email={huguet@cenbg.in2p3.fr}}

\author{J.C. Caillon}{address={Centre d'Etudes Nucl\'{e}aires de Bordeaux-Gradignan, CNRS-IN2P3 \\ Universit\'{e} Bordeaux 1, Le Haut-Vigneau, 33170 Gradignan Cedex, France}}

\author{J. Labarsouque}{address ={Centre d'Etudes Nucl\'{e}aires de Bordeaux-Gradignan, CNRS-IN2P3 \\ Universit\'{e} Bordeaux 1, Le Haut-Vigneau, 33170 Gradignan Cedex, France}}

\begin{abstract}
We have built an effective Walecka-type hadronic Lagrangian in which the hadron masses and
the density dependence of the coupling constants are deduced from the quark dynamics using a Nambu-Jona-Lasinio model. The parameters of this Nambu-Jona-Lasinio model have been determined using the
meson properties in the vacuum but also in the medium through the omega
meson mass in nuclei measured by the TAPS collaboration. Realistic properties of nuclear matter have been obtained.
\end{abstract}




\maketitle

\subsection{Introduction}
One of the most fascinating challenges of nuclear physics is the description of nuclear matter and nuclei starting from Quantum ChromoDynamics (QCD). Even if important progress have been made in QCD calculations on the lattice, such a description is not yet available. Models incorporating the most prominent features of QCD have to be used.

A possibility is to apply the strategy of effective field theories where low energy effective hadronic Lagrangians are obtained by integrating out the degrees of freedom lying above the energy scale considered. This decimation leads to density-dependent masses and couplings in the hadronic Lagrangian \cite{br2}. Such a Lagrangian with density dependent masses and coupling constants determined according to Brown and Rho scaling\cite{b.r} has been proposed by Brown, Song, Min and Rho\cite{son}. The calculation reported in \cite{son}, assuming a scaling law  leading to a decreasing of the vector meson mass of approximately 20\% at saturation, enables a realistic description of bulk properties of nuclear matter.

Recently, new experimental results from TAPS collaboration \cite{tap} and KEK/PS \cite{kek} for the in-medium $\omega$ meson mass would suggest a small decreasing of approximately 10-15\% at saturation density. We have explored the possibility of obtaining a realistic description of bulk properties of nuclear matter in the model of Song et al. \cite{son} with a density dependence of the in-medium $\omega$ mass in accordance with recent experimental indications. In addition, the density dependencies of the lagrangian have been deduced directly from the quark dynamics using an in-medium NJL model, since, despite the lack of confinement, it allows to take into account an important part of quark dynamics with the in-medium chiral symmetry restoration. 

\subsection{Model} 

We use an effective hadronic Lagrangian similar to the Walecka one but with density dependent masses $M^*_N$, $m^*_{\omega}$, $m^*_{\sigma}$ and couplings $g^*_{\sigma NN}$, $g^*_{\omega NN}$. The vacuum values $g_{\omega NN}$, $g_{\sigma NN}$ are the only free-parameters, which will be adjusted at the end of the calculation to reproduce the position of the saturation point. This Lagrangian is treated at the mean-field level for infinite symmetric nuclear matter.
 
The density-dependent couplings and mass parameters are determined directly from the quark dynamics using a two flavor NJL model, which includes two chirally invariant four-quark terms in scalar/pseudoscalar and vector channel, and a scalar/vector eight quark term. At the quark level, the free parameters are the bare quark mass $m_0$, the couplings $g_1$, $g_2$ and $g_3$ and the three momentum cut-off $\Lambda$. The meson masses and couplings to quarks $m^*_{\sigma,\omega}$, $g_{\sigma, \omega qq}$ are determined by solving the Bethe-Salpeter equation in quark-antiquark appropriate channels. We assume that $M_{N}^{*}$ is directly related to the quark condensate,  with the same relation as found in finite-density QCD sum-rule calculations\cite{coh}, and that the quark-meson and nucleon-meson couplings in medium are proportional : 

\begin{equation}
\frac{M_{N}^{*}}{M_{N}}= \frac{\left\langle \overline{q}q\right\rangle 
}{\left\langle \overline{q}q\right\rangle _{0}}, \hspace{5mm} \frac{g^*_{\sigma NN}}{g_{\sigma NN}} = \frac{g_{\sigma qq}^{*}}{g_{\sigma qq}},\hspace{5mm} \frac{g^*_{\omega NN}}{g_{\omega NN}} = \frac{g_{\omega qq}^{*}}{g_{\omega qq}},
\end{equation}

\noindent where $\left\langle \overline{q}q\right\rangle _{0}$ represents
the  quark condensate in vacuum, and $M_N=939$ MeV is the free nucleon mass. More details of the model are available in \cite{huguet}.

\subsection{Results}

At the quark level, for a given value of the cutoff $\Lambda $, we impose to reproduce the pion mass $m_{\pi }=135$ MeV,  the pion decay constant $f_{\pi }=92.4$ MeV and the $\omega $ meson mass $m_{\omega }=782$ MeV in vacuum. We have chosen to take into account the result obtained by the TAPS collaboration\cite{tap} for the $\omega $ meson mass in nuclei, $m_{\omega }^{*}(\rho_{B}=0.6\rho _{0})=722_{-4}^{+4}$ (stat)$_{-5}^{+35}$(syst) MeV in order to constrain the eight-quark term $g_3$ parameter. 

At the hadronic level, the free parameters $g_{\sigma NN}$, $g_{\omega NN}$ are fixed to reproduce the saturation point. For each value of $\Lambda$ considered, in order to probe the description of nuclear matter obtained, we have calculated the effective nucleon mass $m_{N}^{*}/M_N = (M^*_N - \left(g^{*}_{\sigma NN}/m^*_{\sigma}\right)^2 \rho_S)/M_N$ (with $\rho_S$ the nucleon scalar density), the incompressibility parameter $K$ and the slope of the real part of the energy dependence of the nucleon-nucleus optical potential  $U_{0}/M_{N}$ at saturation, which are expected empirically to be respectively of order $0.58-0.6$, $250\pm 50$ MeV, $0.25 - 0.40$. We have plotted on Fig.1 these three quantities as function of the constituent quark mass in vacuum $m$, where the shaded areas correspond to the bound on the empirical value.


As we can see, for $m\approx $ $465-470$ MeV ($\Lambda = 572 \pm 1$ MeV), the three physical quantities $m_{N}^{*}/M_{N}$, $K$ and $U_{0}/M_{N}$ are all in good agreement with the empirical values, and keep reasonable values at least up to $m \approx 500$ MeV. The saturation curve is realistic for density around the saturation point. At higher densities, it is somewhat harder that what is expected, but this is  not surprising since the relationship between $M^*_N$ and $\left\langle \overline{q}q\right\rangle$ should be valid only at low densities. 

\begin{figure}[htb]
\centering
\parbox{16cm}{
  \begin{minipage}{5cm}
      \epsfig{file=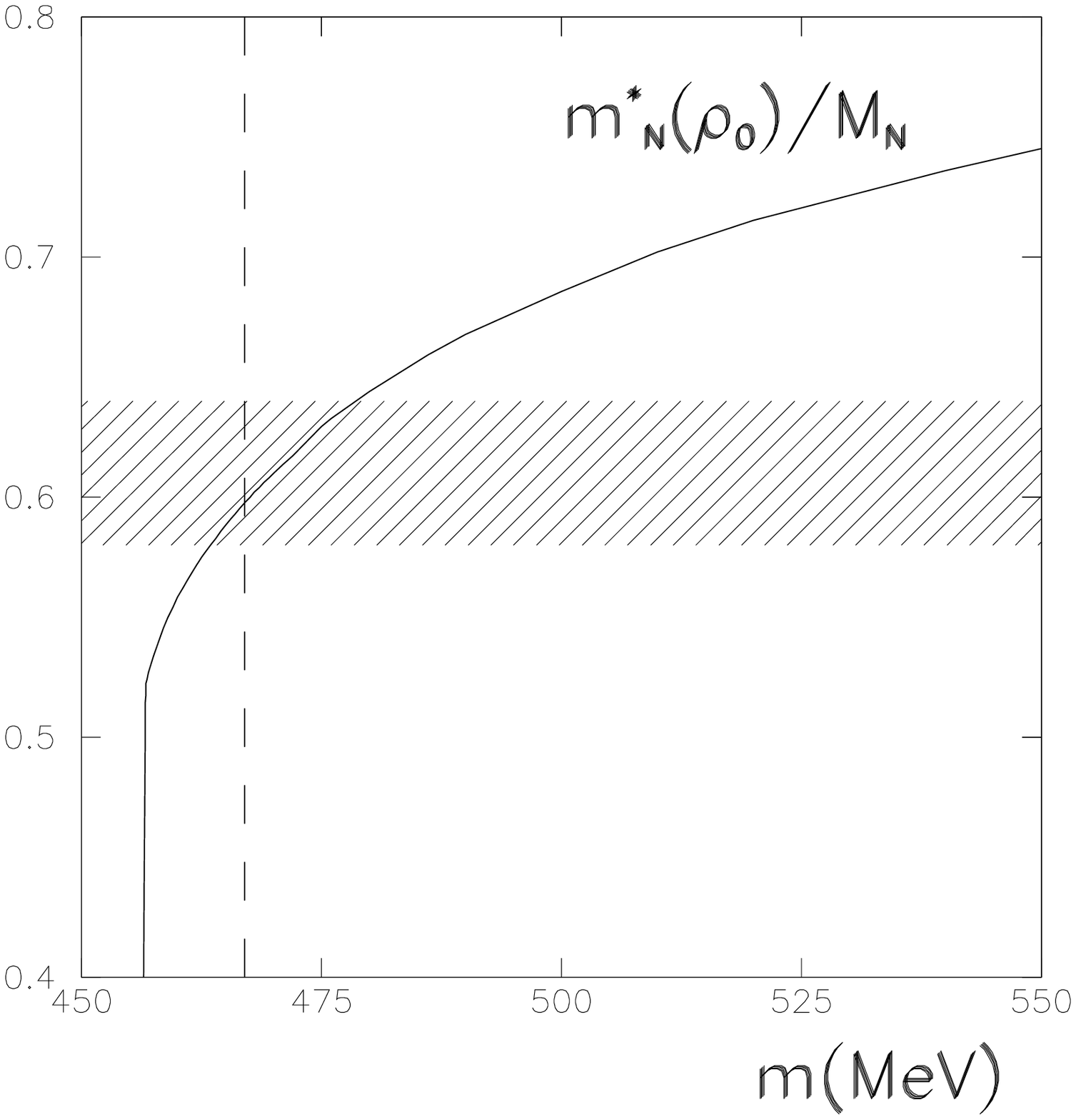,scale=0.26}
  \end{minipage}
  \begin{minipage}{5cm}
      \epsfig{file=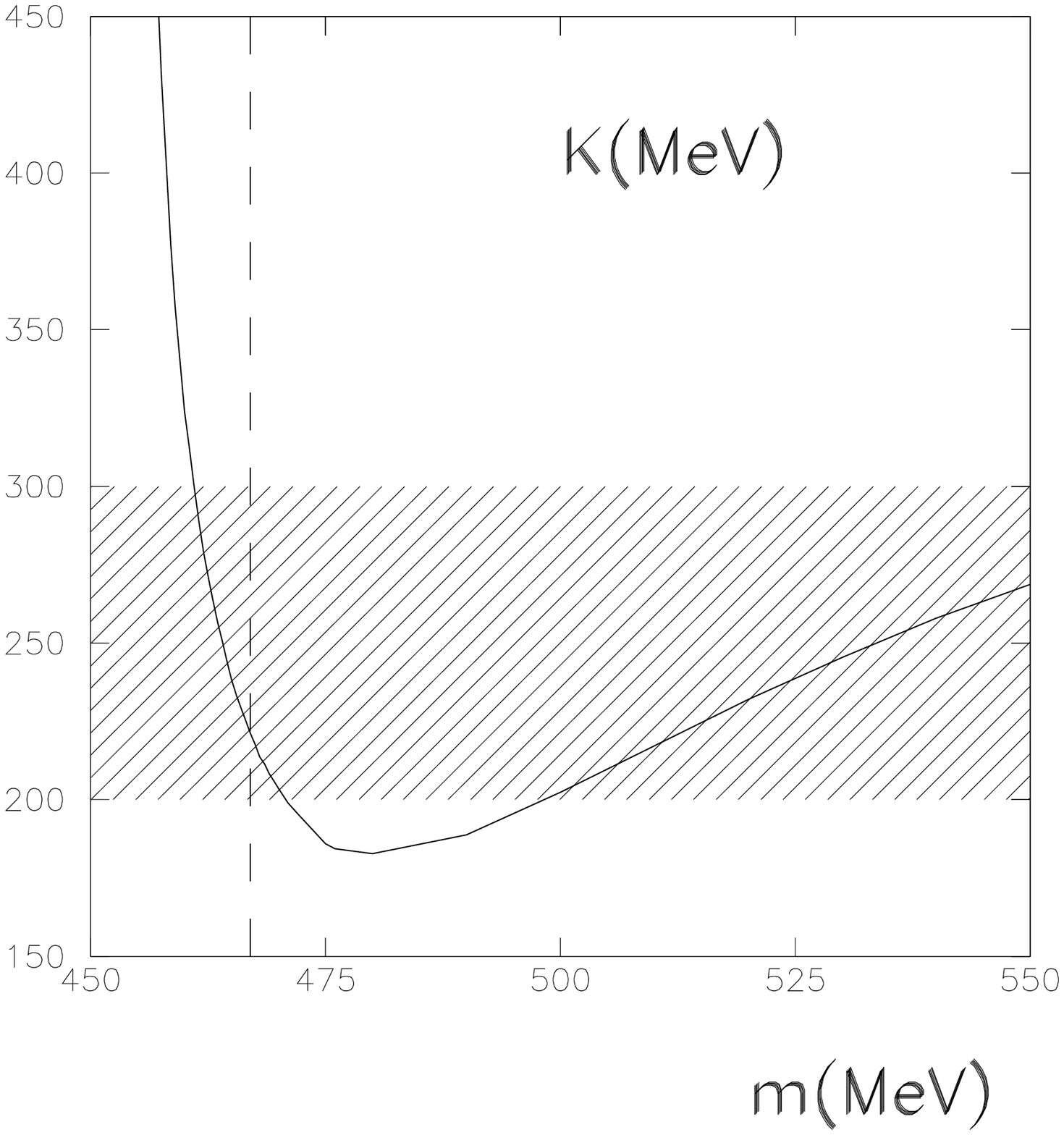,scale=0.26}
  \end{minipage}
  \begin{minipage}{5cm}
      \epsfig{file=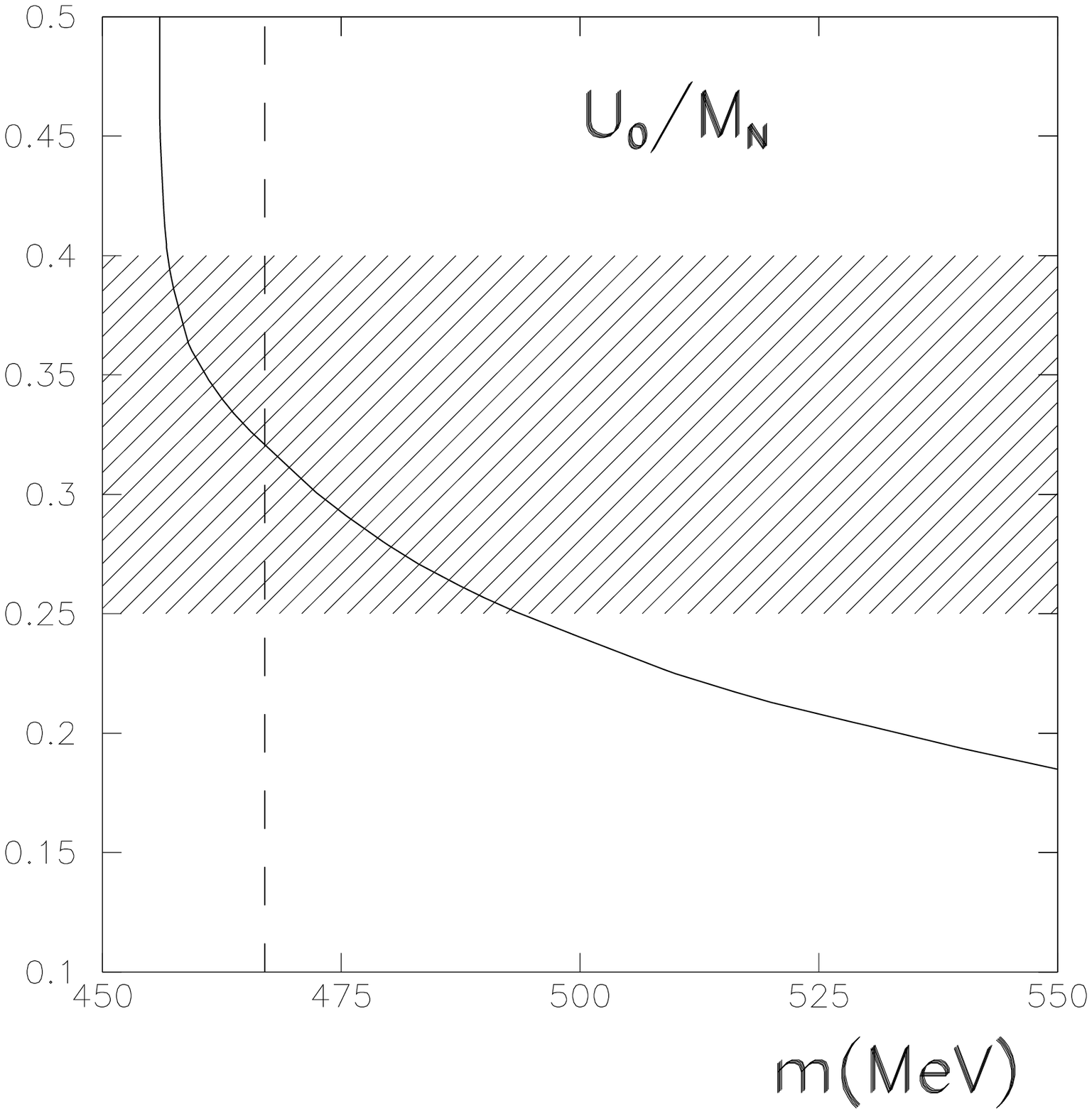,scale=0.26}
  \end{minipage}
  }
  \caption{Saturation properties as function of the constituent quark mass $m$.}
\end{figure}
  
\subsection{Conclusion}
We have investigated the  properties of nuclear matter in a
relativistic mean field model with density-dependent masses and couplings, similar to that used in \cite{son}, in which we have replaced the Brown and Rho scaling by a direct calculation of meson masses and couplings in a NJL quark model.
The NJL model including four and eight quark interaction terms has been constrained to reproduce the recent TAPS result for the in medium $\omega$ meson mass in addition to the vacuum pion and $\omega$ meson properties. At the hadronic level, the two free parameters have been fixed to reproduce the empirical saturation point. 

At the quark level, the eight quark term is essential for obtaining realistic saturation properties. At saturation, the nucleon effective mass, the incompressibility parameter and the slope of the energy dependence of the nucleon-nucleus optical potential obtained are all in good agreement with the empirical data. 

Even if more work is needed in this direction and in a more fundamental one, this result is a very encouraging one since with only a few free-parameters, a realistic description of saturation properties of nuclear matter has been obtained from a hadronic Lagrangian constrained by a quark model which reproduces vacuum and in-medium meson properties.

\end{document}